# 3-mode detection for widening the bandwidth of resonant gravitational wave detectors

L. Baggio,[1] M. Bignotto,[2] M. Bonaldi,[3] M. Cerdonio,[2] L. Conti,[2] P. Falferi,[3,*] N. Liguori,[2] A. Marin,[2] R. Mezzena,[1] A. Ortolan,[4] S. Poggi,[1] G. A. Prodi,[1] F. Salemi,[5] G. Soranzo,[6] L. Taffarello,[6] G. Vedovato,[4] A. Vinante,[1] S. Vitale,[1] and J. P. Zendri[6]

[1] Dipartimento di Fisica, Università di Trento and INFN, Gruppo Collegato di Trento, Sezione di Padova, I-38050, Povo, Trento, Italy

[2] Dipartimento di Fisica, Università di Padova and INFN, Sezione di Padova, Via Marzolo 8, I-35131 Padova, Italy

[3] Istituto di Fotonica e Nanotecnologie, CNR-ITC and INFN, Gruppo Collegato di Trento, Sezione di Padova, I-38050 Povo, Trento, Italy

[4] Laboratori Nazionali di Legnaro, Istituto Nazionale di Fisica Nucleare, 35020 Legnaro, Padova, Italy

[5] Dipartimento di Fisica, Università di Ferrara and INFN, Sezione di Ferrara, I-44100 Ferrara, Italy

[6] Istituto Nazionale di Fisica Nucleare (INFN), Sezione di Padova, Via Marzolo 8, I-35131 Padova, Italy

[*] Corresponding author. Email address: falferi@science.unitn.it

We have implemented a novel scheme of signal readout for resonant gravitational wave detectors. For the first time, a capacitive resonant transducer has been matched to the signal amplifier by means of a tuned high Q electrical resonator. The resulting 3-mode detection scheme widens significantly the bandwidth of the detector. We present here the results achieved by this signal readout equipped with a two-stage SQUID amplifier. Once installed on the AURIGA detector, the one-sided spectral sensitivity obtained with the detector operated at 4.5 K is better than $10^{-20}$ $Hz^{-1/2}$ over 110 Hz and in good agreement with the expectations.

Along with the peak sensitivity an important parameter of a resonant gravitational wave detector is its bandwidth. In addition to the obvious advantage of making the detector more sensitive to the short bursts, a wider bandwidth would allow, for instance, to resolve details of the signal emitted during a supernova gravitational collapse or the merger of compact binaries[1]. Moreover, a wider bandwidth reduces the uncertainty in the burst arrival time[2] and consequently, with a detector network, permits a more precise source location and a higher efficiency of spurious events rejection[3]. The introduction of a mechanically resonant transducer, a standard practice in actual resonant detectors, has greatly improved the coupling between the bar and the amplifier but the bandwidth is intrinsically limited,[4] and in practice, according to the Full Width Half Maximum (FWHM) definition applied to the two minima of the $S_{hh}$ strain noise spectra, values of a few Hz have been achieved[5]. The use of multimode resonant transducers should permit further improvements of the detector bandwidth[6]. This approach has been studied[7] in-depth and a few 2-mode transducer prototypes have been realized[8] or are under development[9] to obtain 3-mode operation of the resonant mass detectors.

This letter describes how a wider detection bandwidth can be obtained with an alternative 2-mode transduction system in which the resonant amplification is realized by means of a resonant mechanical mode plus a resonant electrical matching network. It also describes the key tests performed on the components of the transduction system in order to verify the achievement of the requirements set by analysis of the detector model.

Figure 1 shows the electromechanical scheme of a cryogenic detector with a resonant capacitive transducer read by a SQUID amplifier. The matching transformer couples the output impedance of the transducer (a capacitance of a few nF) to the input impedance of the SQUID (a small inductance of the order of µH). The current in the SQUID input coil $L_i$ is converted by means of the coupling $M_i$ in a magnetic flux at the SQUID loop and then in the SQUID output voltage. In addition to the two mechanical modes of bar and transducer, there is then a third electrical mode formed by the transducer capacitance $C_T$ (the effect of the decoupling capacitor $C_d \gg C_T$ is negligible) and by the inductance L of the primary coil of the matching transformer. With practical values of L and $C_T$, it is possible to tune the electrical mode to the mechanical modes.

In the equations of motion of the system,[10] the only considered noise sources are the thermal noise due to the losses of bar, transducer, and electrical mode and the current (additive) and voltage (back action) noise of the SQUID amplifier. From these equations we derive with numerical computations the equivalent input one-sided strain noise $S_{hh}^{1/2}$ which usually expresses the sensitivity of gravitational wave detectors. To do this we consider the typical operating parameters of the AURIGA detector: bar mass 2300 kg, optimized effective mass of the transducer 3.5 kg, bar and transducer resonance frequency 897 Hz, bar quality factor $6.4 \times 10^6$, transducer quality factor $8 \times 10^5$, transducer electric bias field $7.5 \times 10^6$ V/m, operating temperature 2K. For the two stage SQUID additive and back action noise spectra, the conservative values of $S_{ii} = 3.2 \times 10^{-26}$ A$^2$/Hz and $S_{vv} = 3.2 \times 10^{-30}$ V$^2$/Hz have been chosen[11].

The sensitivity spectra $S_{hh}$ obtained from the numerical computations show that the detector sensitivity to the short bursts, expressed in terms of the effective temperature $T_{eff}$ ($T_{eff} = \Delta E_{min}/k_B$ where $\Delta E_{min}$ is the minimum detectable energy in the detector), can improve up to a factor 3 when the electrical mode with a quality factor $\approx 10^6$ is tuned to the mechanical modes and that a sensitivity $S_{hh}^{1/2}$ better than $10^{-20}$ Hz$^{-1/2}$ over a bandwidth of about 100 Hz can be achieved. The effect of the electrical mode tuning on the effective temperature is even more impressive if a single stage SQUID amplifier with a typical energy resolution of 10000 $\hbar$ is considered. In this case $T_{eff}$ improves by a factor of about 10.

The electrical mode is considered tuned when its frequency equals that of the uncoupled mechanical modes but we find that a mismatch up to 10 Hz does not significantly affect the detector bandwidth. The simulations with detuned electrical mode have been obtained with the electrical mode at 2000 Hz. The quality factor of the electrical mode has a critical importance for the detector sensitivity: the tuning is considerably advantageous only if its quality factor is at least of order $10^6$. For our operating parameters any further increase of quality factor produces a negligible effect. In previous capacitive transducers read by a SQUID amplifier, the electrical mode has been kept detuned because the techniques necessary for the achievement of electrical quality factor of order $10^6$ were not yet known and a Q of order $10^4$ would have spoiled the mechanical mode quality factors with a consequent reduction of the detector sensitivity. From the experience gained from experimental investigation of high Q electrical resonators (Q~$10^6$), first developed to study the stability problem of the resonator-SQUID system and to characterize the SQUID noise sources[12], it has turned out that the conditions for a more advantageous matching could be achieved.

To check the feasibility of this 3-mode detection scheme, a new readout for the AURIGA detector has been realized and tested in the cryogenic Transducer Test Facility[13] at the INFN Legnaro Laboratories. The readout is composed of a resonant mushroom capacitive transducer[14] with capacitance 9.3 nF and an impedance matching transformer, housed in a superconducting box. The box is attached to the transducer by means of a suspension that decouples the two elements in the frequency region around 1 kHz in order to prevent the mechanical dissipation in the box from reducing the transducer and bar quality factors. The electrical scheme of the readout is shown in Fig. 2(a). The transformer box houses also, in separate compartments, the two units which form the two-stage SQUID amplifier[11]. In this device the signal of the first SQUID (the sensor SQUID) is amplified by a second SQUID (the amplifier SQUID) in order to make negligible the noise contribution of the room temperature electronics and achieve the intrinsic noise of the sensor SQUID which is expected to scale with the temperature. In addition to these characteristics the employed SQUID amplifier has three peculiar features that make it suitable to the particular input load constituted by the detector readout:

(1) When a SQUID is strongly coupled to a high Q input load, the real part of its dynamic input impedance, which can assume negative values, can drive the whole system to instability. An RC filter placed between the input coil and the feedback line, is used as a cold damping network[15] to avoid negative Q instabilities without adding noise (Fig. 2(a)). Besides avoiding instabilities, the damping network can reduce the quality factors of the detector modes to typical values of order 100÷1000. This allows faster detector response times and simpler data analysis.

(2) The intrinsic quality factor of the modes needs to be known to check if the noise at the modes is thermal. Unfortunately, the SQUID dynamic input impedance and the cold damping network alter the quality factor in a way that cannot be evaluated with sufficient precision. For this reason, in addition to the transformer secondary coil, a small pick-up coil is weakly coupled (Fig. 2(a)) to the transformer primary coil and is part of the superconducting transformer placed between the sensor SQUID and the amplifier SQUID. By biasing the sensor SQUID with a much higher than optimal current (a few mA instead of 30 µA), this is made insensitive and, in accordance with the RSJ model[16] of the Josephson junction with shunting resistance, it is equivalent to a superconducting loop with inductance $L_{SQ}$ interrupted by two shunt resistances $R_S$. Given the couplings and the values of $L_{SQ}$ and $R_S$ specified by the manufacturer[17] ($L_{SQ}$ = 80 pH and $R_S$ = 2±1 Ω), the sensor SQUID with high bias current has negligible effect on the quality factor of the modes (in particular the electrical mode which is more sensitive to the coupling with the SQUID) up to quality factors of 2-3 million. In these operating conditions the amplifier SQUID, operated in closed loop mode, can measure with the weakly coupled pick-up the so-called "intrinsic quality factor" which is determined not only by the losses in the resonator but also by possible losses in the sensor SQUID coupling circuit and in the sensor SQUID loop itself.

(3) Another peculiar characteristic of the employed two-stage SQUID is the calibration coil integrated in the sensor SQUID (Fig. 2(a)). The calibration coil is coupled to a pick-up coil of negligible inductance in series with the SQUID input coil $L_i$ and the secondary coil $L_S$. From the transfer function between the SQUID output signal and the calibration coil input signal, it is possible to evaluate the equivalent impedance $Z_{eq}$ at the input of the SQUID seen as current amplifier.

The Teflon decoupling capacitor and the cryogenic switch (Fig. 1) are housed in separate cases connected to the transformer box by means of a suspension designed to have no resonances around 1 kHz. The main purpose of the cryogenic switch is to avoid that the electrical quality factor is spoiled by the losses of the stray capacitance of the charging cable, in particular of the part at room temperature. The resonance frequency $\omega_0/2\pi$ of the resultant electrical mode is 898 Hz. The entire readout, with its mechanical suspension system, is housed in the vacuum chamber of the Transducer Test Facility and cooled to temperatures between 1.3 and 4.2 K in a liquid helium bath measured by a germanium thermometer located on the transformer box.

We stress that the advantage of this detection scheme depends only on the achievement of a quality factor of order $10^6$ and the thermal noise level in the LC circuit, formed by the transducer and the matching transformer primary. For this reason, the transducer has been kept uncharged in the measurements described in the following and, as a result, only the electrical mode is present.

Fig. 2(b) shows the noise model under these conditions. $S_{vv}$ and $S_{ii}$ are the one-sided spectral densities of the back action noise $V_n$ and additive noise $I_n$ of the two-stage SQUID amplifier, which are assumed to be uncorrelated. The SQUID noise theory[18] predicts $S_{ii}=16k_BTL_{SQ}^2/R_SM_i^2$ and $S_{vv}=11k_BT\omega_0^2M_i^2/R_S$. The thermal noise source $V_{Th}$ with spectral density $4k_BTr$ is due to the intrinsic losses of the resonator-SQUID system which determine the intrinsic quality factor as described above. C represents the series of the transducer capacitance and the decoupling capacitance. The real part of the SQUID dynamic input impedance and the cold damping network are represented by a noise-free resistor $r_c$. The impedance $Z_{eq}$ represents the SQUID input load and is defined by $Z_{eq}=j\omega M_{cal}(I_{cal}/I_{in})$ where $I_{cal}$ is the current in the calibration coil and $I_{in}$ is the input current of the SQUID amplifier. From the above described noise model and from the definition of $Z_{eq}$ one can show that the expected current noise spectral density at the SQUID input is

$$S_I(\omega) = 4k_B T \left(\frac{Q_a}{Q}\right) \Re\left\{\frac{1}{Z_{eq}(\omega)}\right\} + S_{vv}(\omega)\frac{1}{|Z_{eq}(\omega)|^2} + S_{ii}(\omega) \qquad (1)$$

where Q is the intrinsic quality factor measured with the sensor SQUID biased with high current and $Q_a$ is the apparent quality factor measured when the two-stage SQUID is operated in closed loop mode and the system is stabilized by the cold damping network. The three terms on the right-hand side of Eq. (1) are respectively due to the resonator thermal noise, the SQUID back action noise, and the SQUID additive noise. Given the relatively high apparent quality factor (typical values are of order 100), the contribution of $S_{ii}$ is negligible near the resonance frequency and both the thermal and back action terms have a functional form approximated by Lorentzian curves of the type $1/[(1-(\omega_0/\omega)^2)^2+(\omega_0/\omega Q_a)^2]$.

From the output noise spectra of the SQUID and its response to the signal fed to the calibration coil it has been possible to demonstrate that the noise associated with the electrical mode of the capacitive readout is in good agreement with the noise model discussed above, that is with Eq. (1), in the whole temperature range 1.3 – 4.2 K. In particular the electrical mode equivalent temperature[12] $T_m$ which express the energy of the current fluctuations in the LC circuit at the electrical mode frequency, is proportional to the operating temperature T: the best linear fit of the data shown in Fig. 3 is $T_m=(1.07\pm0.05)T+(0.1\pm0.1)$ K. The slight discrepancy in respect to unity for the slope is consistent with the contribution due to the SQUID back action noise $T_{ba}=0.06T$ expected from the theory. The measured intrinsic quality factors of about $5\times10^5$ resulted sufficiently close to the required value of $10^6$ and weakly dependent on the operating temperature. The difference with respect to the expected value is probably due to dissipative elements in the sensor SQUID loop[19]. The SQUID broadband flux noise measured between 3.5 and 4 kHz, well above the resonance peak, scales with the temperature. The best fit is $S_\Phi = (3.8 \pm 0.1) \times 10^{-13} T$ ($\Phi_0^2$/Hz).

Thanks to these key results and after the achievement of the target specifications as regards mechanical quality factor and bias electric field, the entire readout assembly was installed on the AURIGA detector. The AURIGA detector started a run[20] at 4.5 K in December 2003. Figure 4 shows a recent one-sided strain noise spectrum $S_{hh}^{1/2}$ which is better than $10^{-20}$ Hz$^{-1/2}$ over a bandwidth of about 110 Hz and in good agreement with the expectation. According to the FWHM definition applied to the two minima of the $S_{hh}$ spectrum, the bandwidth is 26 Hz. The effective temperature is 320 µK.

In conclusion, we have shown that a resonant capacitive readout, which takes advantage of the tuning of the electrical mode to the mechanical modes, is feasible and improves quite significantly the detector bandwidth. Further improvements could be achieved by cooling the entire detector (bar and readout) to ~100 mK and increasing the bias electric field in the transducer by a factor 2.5. In this case the expected spectral sensitivity is better than $10^{-21}$ Hz$^{-1/2}$ over 130 Hz and $T_{eff}$=7 µK.

This work was supported in part by Grant No. 9902193538 MURST-COFIN '99.

FIG. 1. Electromechanical scheme of the AURIGA detector. L=7.89 H and $L_S$=3.48 µH and their coupling constant is k=0.86. The primary coil inductance, reduced by the coupling to the SQUID, is $L_r$=L-M²/($L_i$+$L_S$)=3.80 H.

FIG. 2. (a) Schematic circuit diagram of the two-stage SQUID coupled to the electrical resonator that represents the uncharged capacitive readout. $L_i$= 1.48 µH, $M_i$= 9.47 nH, R1=470 kΩ, R2=470 kΩ, C1=2.5 nF. (b) Noise model of the resonator – SQUID system.

FIG. 3. Electrical mode equivalent temperature as a function of the operation temperature with the best linear fit $T_m$=(1.07±0.05)T+(0.1±0.1) K.

FIG. 4. Strain noise spectrum $S_{hh}^{1/2}$ of the AURIGA detector (in grey) and that expected from the noise model (in black).

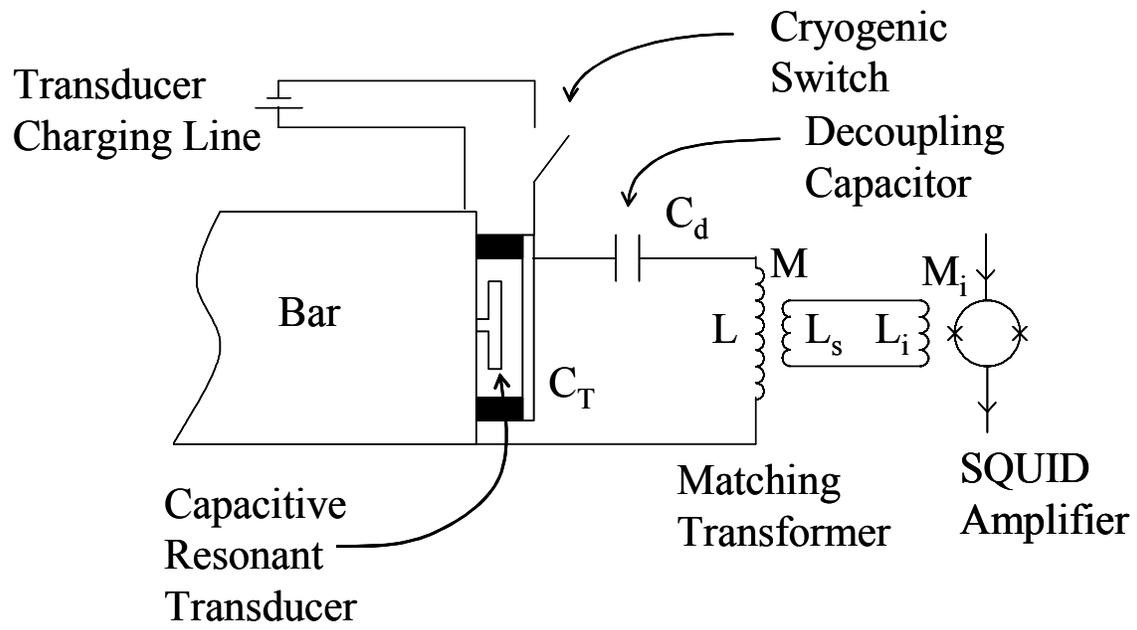

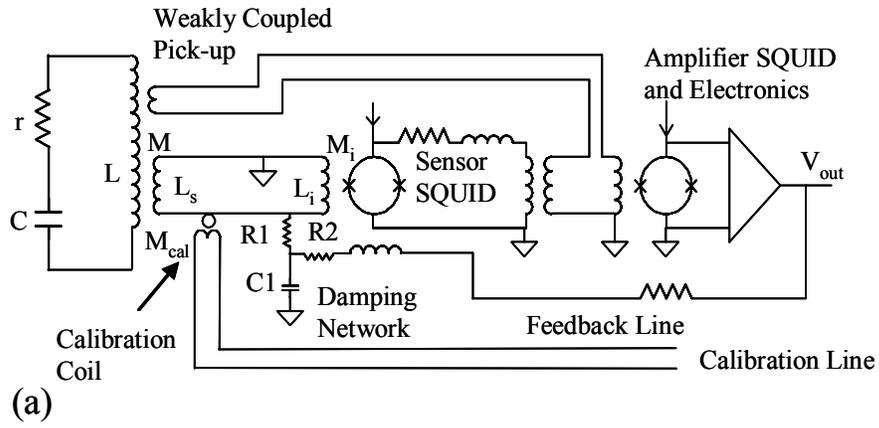

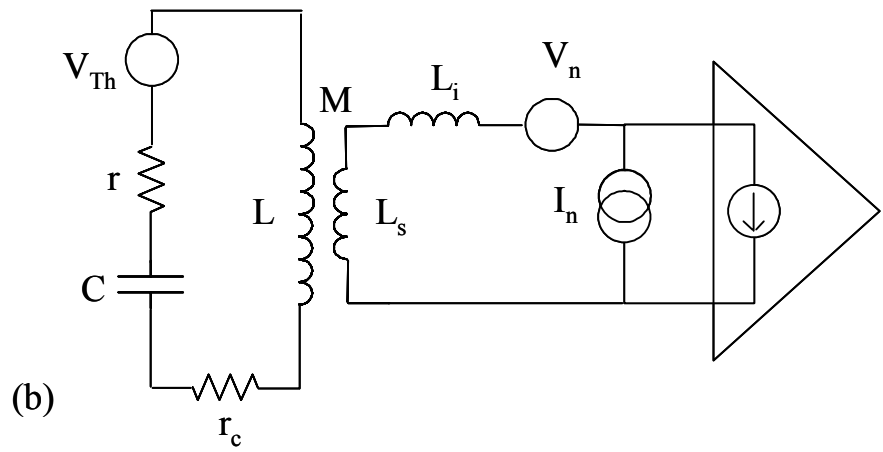

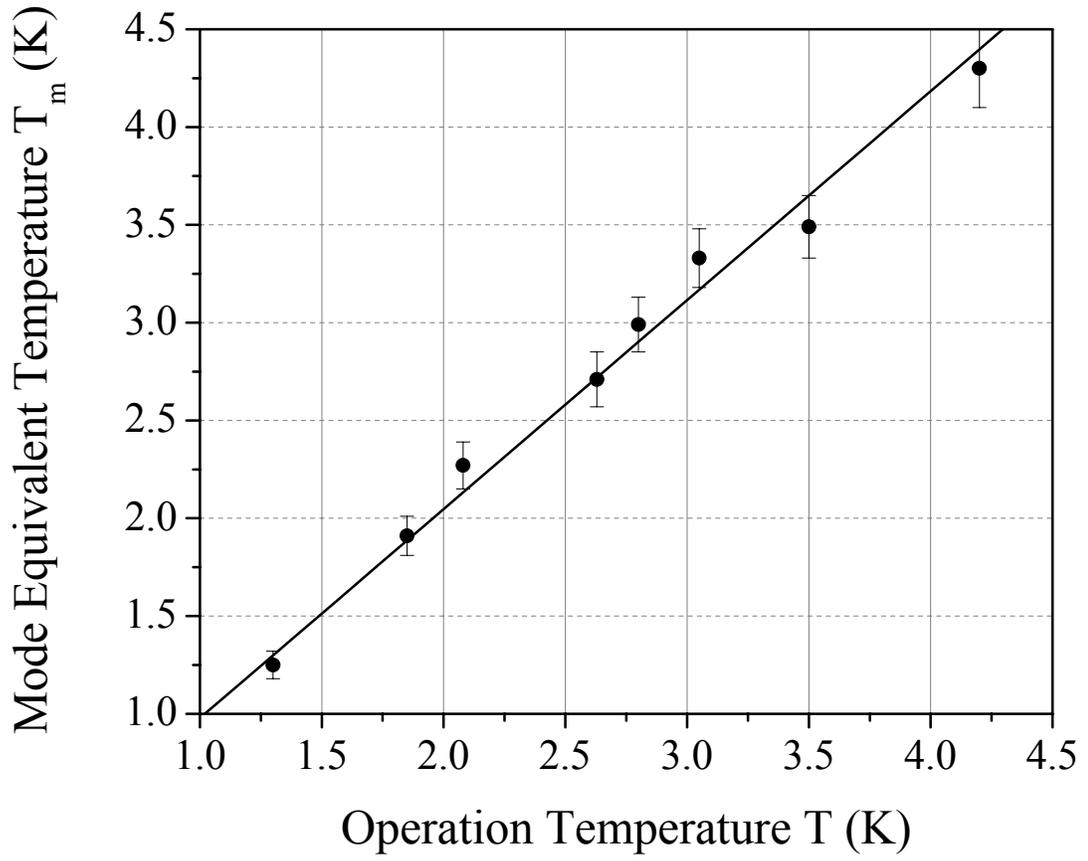

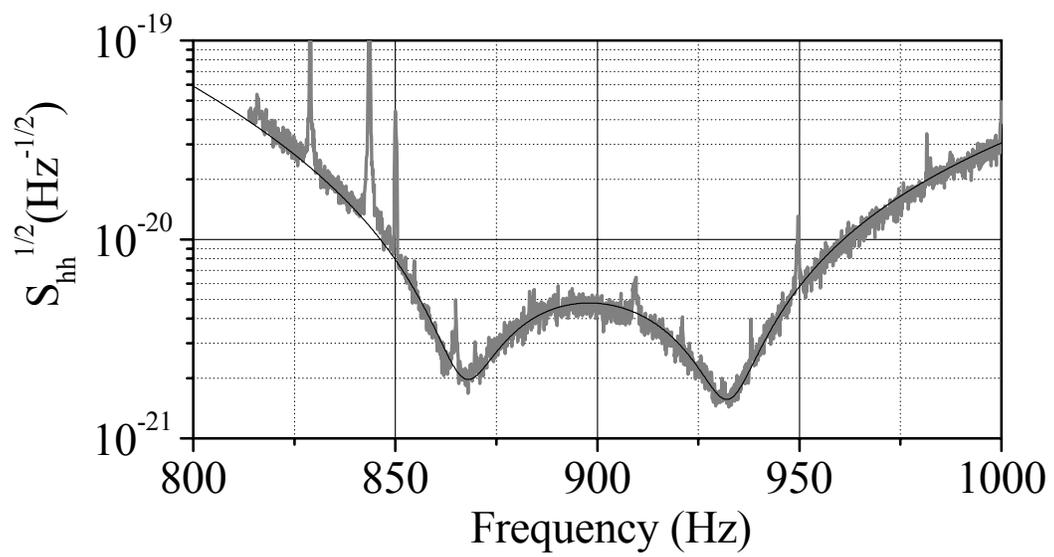

---

[1] C. Cutler and K. S. Thorne, in Proceedings of the GR16 Conference on General Relativity and Gravitation, Durban (South Africa), 2001, edited by N. T. Bishop and S. D. Maharaj (World Scientific, Singapore, 2002).

[2] V. Crivelli-Visconti *et al.*, Phys. Rev. D **57**, 2045 (1998).

[3] L. Baggio *et al.*, Class. Quant. Grav. **19**, 1541 (2002).

[4] H. J. Paik, in Proceedings of the 1st Edoardo Amaldi Conference on Gravitational Wave Experiments, Frascati (Rome), 1994, edited by E. Coccia, G. Pizzella, and F. Ronga (World Scientific, Singapore, 1995), p. 201.

[5] P. Astone *et al.*, Phys. Rev. Lett. **91**, 111101 (2003); E. Mauceli *et al.*, Phys. Rev. D **54**, 1264 (1997); D. G. Blair *et al.*, Phys. Rev. Lett. **74**, 1908 (1995); G. A. Prodi *et al.*, in Proceedings of the 2nd Edoardo Amaldi Conference on Gravitational Waves Experiments, CERN, Geneva, Switzerland, 1997, edited by E. Coccia, G. Veneziano, and G. Pizzella, (Word Scientific, Singapore, 1998), p. 148.

[6] J.-P. Richard, Phys. Rev. Lett. **52**, 165 (1984).

[7] J. C. Price, Phys. Rev. D **36**, 3555 (1987).

[8] M. Bassan, F. Buratti, and I. Modena, Nuovo Cimento **109 B**, 897 (1994); M. P. McHugh *et al.*, in Proceedings of the 2nd Edoardo Amaldi Conference on Gravitational Waves Experiments, CERN, Geneva, Switzerland, 1997, edited by E. Coccia, G. Veneziano, and G. Pizzella, (Word Scientific, Singapore, 1998), p. 413.

[9] Two-mode transducers are under development by the groups of the ALLEGRO detector, MINIGRAIL detector, and MARIO SCHENBERG detector

[10] G. V. Pallottino and G. Pizzella, Nuovo Cimento **4C**, 237 (1981); P. F. Michelson and R. C. Taber, J. Appl. Phys. **52**, 4313 (1981).

[11] R. Mezzena *et al.*, Rev. Sci. Instrum. **72**, 3694 (2001); A. Vinante *et al.*, Appl. Phys. Lett. **79**, 2597 (2001).

[12] P. Falferi *et al.*, J. Low Temp. Phys. **123**, 275 (2001).

[13] M. Bignotto, Laurea thesis, Univ. of Padova, 1999-2000, unpublished (in Italian)

[14] P. Rapagnani, Nuovo Cimento **5 C**, 385 (1982).

[15] A. Vinante *et al.*, Physica C **368**, 176 (2002).

[16] W. C. Stewart, Appl. Phys. Lett. **12**, 277 (1968).

[17] Quantum Design, 11578 Sorrento Valley Road, Suite 30, San Diego, CA 92121-1311, USA.

[18] J. Clarke, C. Tesche, and R. P. Giffard, J. Low Temp. Phys. **37**, 405 (1979).

[19] P. Falferi *et al.*, Appl. Phys. Lett. **82**, 931 (2003).

[20] See for up to date results: http://www.auriga.lnl.infn.it/